# Partitionable Diffractive Neural Networks for Multifunctional Optical Operations

Yudong Tian,[1,†] Haifeng Xu,[2,†] Yuqing Liu,[1] Xiangyu Zhao,[1] Jierong Cheng,[2,*] and Chongzhao Wu[1,*]

[1]Center for Biophotonics, Institute of Medical Robotics, School of Biomedical Engineering, Shanghai Jiao Tong University, Shanghai, China
[2] Institute of Modern Optics, Nankai University, Tianjin, China
[†]These authors contributed equally to this work.
* Corresponding author: chengjr@nankai.edu.cn (Jierong Cheng) and czwu@sjtu.edu.cn (Chongzhao Wu)

Diffractive neural network (DNN), which can perform machine learning tasks based on the light propagation and diffraction, has recently emerged as a promising optical computing paradigm due to its high parallel processing speed and low power consumption nature. However, existing diffractive network architectures face challenges in implementing functional reconfiguration. Once a diffractive neural network is fabricated, its functionality is fixed. Deploying such systems for different tasks typically requires reconstructing the entire physical setup, which significantly compromises hardware efficiency in practical applications. In this work, we propose the multifunctional partitionable diffractive neural networks (PDNNs) that can generate networks with additional capabilities by stacking multiple sub-modules with independent functions in the horizontal direction. Each submodule functions as an independent diffractive network capable of performing specific imaging or classification tasks. When these submodules are combined, they can form a new network with additional functionalities. Moreover, assembling these submodules in different configurations enables structures with diverse functions. This powerful PDNN framework demonstrates remarkable advantages in flexibility and reconfigurability for multitask operations, opening a new pathway for realizing multifunctional and integrated optical artificial intelligence systems.

## 1. Introduction

Deep learning, as a representative paradigm in the field of artificial intelligence (AI), has achieved remarkable progress over the past decades, driving breakthroughs in fields such as computer vision [1,2], natural language processing [3], and autonomous systems [4,5]. The impressive performance of modern deep learning models largely relies on their ever-increasing model size, depth, and parameter complexity, which in turn demand massive computational resources and energy consumption. Conventional electronic processors, including CPUs, GPUs, and specialized accelerators such as TPUs, have pushed the boundaries of digital computing by offering parallelism and optimized architectures for training and inference of deep learning models. However, with the rapid scaling of AI models, these hardware platforms face significant bottlenecks in terms of computation speed, memory bandwidth, and most critically, power efficiency. For example, training state-of-the-art neural networks often requires hundreds of GPU-hours, while the deployment of such models in resource-constrained environments is severely limited by heat dissipation and energy costs. The growing disparity between computational demands and hardware limitations has motivated the exploration of alternative computing paradigms, among which optical neural networks (ONNs) [6–13] have emerged as a promising solution that leverages the intrinsic parallelism, ultrafast propagation, and low energy consumption of light to accelerate deep learning tasks.

All-optical diffractive neural networks (DNNs) [14], as a promising ONN architecture, have attracted particular attention, as they utilize the principles of light diffraction to perform computation through passive optical elements, enabling ultrafast, energy-efficient, and scalable machine learning systems. In a typical diffractive neural network, a series of diffractive layers designed using deep learning algorithms (such as error backpropagation and gradient descent) collectively implement the desired optical field transformation. Each layer is composed of thousands of diffractive units—subwavelength pillars fabricated through techniques such as 3D printing or photolithography—that modulate the phase or amplitude of the optical field. These units are interconnected with those in adjacent layers through light diffraction, and this ultrahigh density and massive parallelism endow the network with superior speed and throughput compared to other forms of ONNs. Recent studies have demonstrated the potential of DNNs in diverse applications, including object classification, broadband pulse shaping [15], image encryption [16,17], optical imaging [18–22], and photonic computing [23,24], among others [25–28], highlighting their advantages of parallel computation and hardware efficiency. However, conventional DNN architectures are primarily designed with a single function and inherently lack flexibility for handling multiple tasks simultaneously. Once trained and fabricated, the diffractive layers are fixed, making it difficult to reconfigure the network for different functions. If extra tasks need to be performed, the entire DNN system has to be rebuilt, which consumes a lot of computing resources and reconstruction costs. Although there have been some studies on the reconfigurable architectures for DNNs, such as pluggable multitask diffractive neural networks [29], hardware-software co-design architecture [30], modular diffractive neural networks [31], among others [32], these methods typically introduced additional costs or required complex experimental setups. More importantly, they primarily focus on vertical stacking of diffractive layers while overlooking the potential for horizontal expansion.

In this work, we proposed a partitionable diffractive neural network (PDNN) architecture constructed by horizontally concatenating multiple independent sub-DNNs, thereby enabling multifunctional optical computation. Each subnetwork in the PDNN functions as an independent module capable of handling distinct machine learning tasks. Meanwhile, these modules can be horizontally assembled to form new diffractive networks with additional functionalities distinct from those of the individual modules. Furthermore, networks assembled in different ways (e.g., by rotating subnetworks at various angles before concatenation) can also be employed to perform different tasks, thereby endowing the PDNN architecture with remarkable flexibility and a high degree of task integration. As proof-of-concept demonstrations, we designed and fabricated two single-layer PDNNs for different application scenarios, each consisting of four submodules. In the first scenario, each submodule is capable of generating terahertz holographic images of the digits "0", "1", "2", "5", while the newly formed network obtained by concatenating these submodules projects the holographic image of digit "7". In the second scenario, each submodule produces holographic patterns of the letters "S", "J", "T", and "U", whereas the combined network exhibits image classification capability and can be employed for MNIST handwritten digit [33] recognition. All experimental measurements closely match our expected results, demonstrating the outstanding reconfigurability of PDNN in both holographic imaging and object classification tasks. In addition, we numerically demonstrated that combining subnetworks after rotating them at different angles can generate new networks with distinct additional functions, thereby revealing the potential of PDNNs for large-scale multi-task integration. The proposed PDNN framework, for the first time, demonstrates the lateral expansion capability of diffractive neural networks. As a general design paradigm, PDNNs can be seamlessly integrated with other multi-task frameworks to further enhance its performance. We believe that this simple yet powerful framework represents a significant step toward practical, multifunctional all-optical artificial intelligence systems capable of real-time multi-task processing.

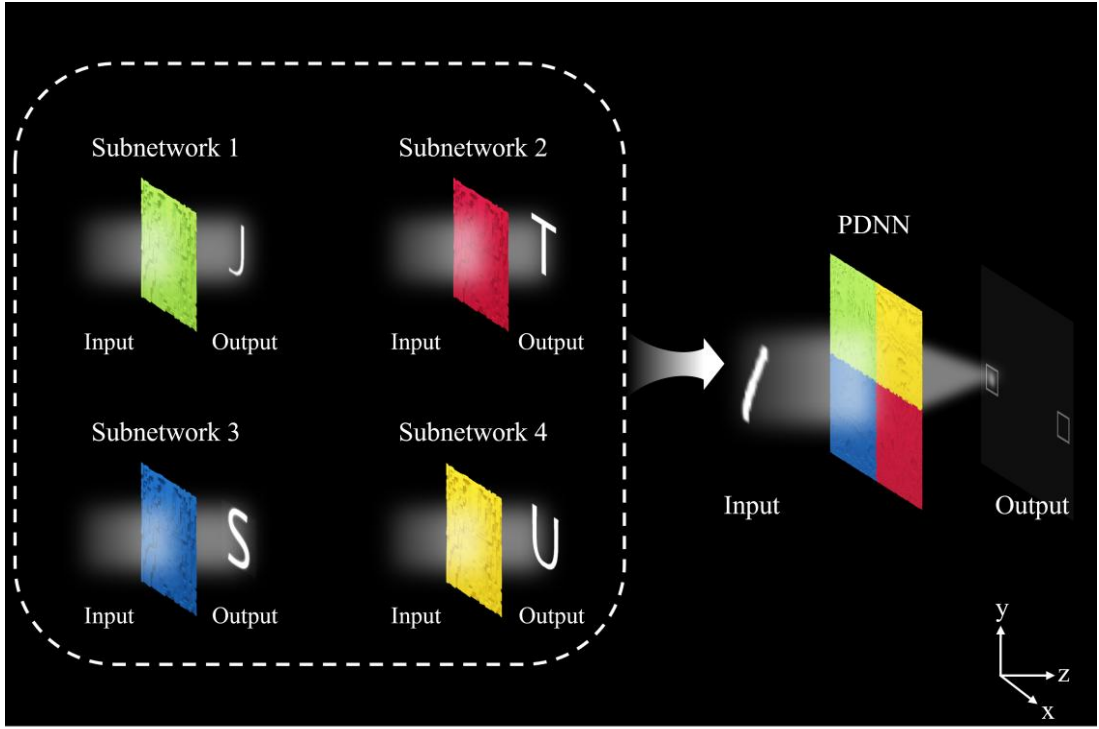

Fig. 1. Schematic of the multifunctional partitionable diffractive neural network (PDNN). The PDNN consists of four subnetworks, where each subnetwork can function as an independent system to perform a specific machine learning task. Meanwhile, these subnetworks can also be horizontally assembled to form new networks with additional functionalities.

## 2. Methods and Materials

## 2.1 Numerical model of PDNN

Fig. 1 illustrates the schematic of the proposed PDNN, which consists of four sub-networks. When operating independently, each sub-network possesses its own input layer, hidden layer and output layer, and can function as a standalone neural network. Meanwhile, the four subnetworks can be horizontally concatenated to form a new network, which is capable of performing machine learning tasks that differ from those of the individual subnetworks.

Similar to traditional DNNs, the PDNN system is trained based on the optical diffraction theory. According to the Rayleigh-Sommerfeld diffraction theory [34], each single neuron of diffractive

layers can be considered as a secondary wave source composed of the following optical mode:

$$w_i^l(x_i, y_i, z_i) = \frac{z - z_i}{r_i^2}\left(\frac{1}{2\pi r_i} + \frac{1}{j\lambda}\right) \times exp\left(\frac{j 2\pi r_i}{\lambda}\right) \quad (1)$$

where $l$ represents the $l^{th}$ diffractive layer of the network, $i$ represents the $i^{th}$ neuron located at ($x_i$, $y_i$, $z_i$) of layer $l$, $\lambda$ is the illumination wavelength, $r_i = \sqrt{(x - x_i)^2 + (y - y_i)^2 + (z - z_i)^2}$ is the propagation distance of this diffractive process and $j = \sqrt{-1}$ is the imaginary number. The amplitude and phase of the modulated optical field are determined by the transmission coefficient (t) of the neuron and its input wave, which can be expressed as:

$$n_i^l(x_i, y_i, z_i) = w_i^l(x_i, y_i, z_i) \cdot t_i^l(x_i, y_i, z_i) \cdot \sum_j n_j^{l-1}(x_i, y_i, z_i) \quad (2)$$

where $\sum_j n_j^{l-1}(x_i, y_i, z_i)$ represents the total incident waves to the $i^{th}$ neuron of layer $l$, and $t_i^l(x_i, y_i, z_i) = a_i^l(x_i, y_i, z_i) \cdot exp(\varphi_i^l(x_i, y_i, z_i))$ is the neuron's transmission coefficient. In this study, since we used phase-only diffractive layers, the amplitude term $a_i^l(x_i, y_i, z_i)$ is assumed to be a constant, ideally 1, ignoring the optical losses during propagation and only the phase term $\varphi_i^l(x_i, y_i, z_i)$ is used as a learnable parameter to construct the network.

Assuming that the network is composed of $M$ diffractive layers, then the measured diffraction field intensity at the output plane can be expressed as:

$$I^{M+1} = |U^{M+1}|^2 \quad (3)$$

where $U^{M+1}$ is the light field distribution at the output plane.

The loss function is defined as the mean squared error (MSE) between the target pattern $T$ and the normalized intensity of the resulting optical field $I^{M+1}$:

$$L_{loss} = \frac{1}{N} \sum_1^N \left| \frac{I^{M+1}}{max(I^{M+1})} - T \right|^2 \quad (4)$$

where N represents the total number of pixels on the output plane.

During the computer-based training stage, the complete network parameters of the PDNN are represented by a matrix G with dimensions of (M, N, N), where M denotes the number of diffractive layers and N represents the number of neurons in each layer. To facilitate training, we partition the entire system into four independently operable subnetworks, denoted as $\varphi_k, k = 1,2,3,4$, each corresponding to one quadrant of the full network. Without loss of generality, here we can define the parameters of each sub-network as:

$$\varphi_1 = G[:, 0:N/2, 0:N/2] \quad (5)$$
$$\varphi_2 = G[:, 0:N/2, N/2:N] \quad (6)$$
$$\varphi_3 = G[:, N/2:N, 0:N/2] \quad (7)$$
$$\varphi_4 = G[:, N/2:N, N/2:N] \quad (8)$$

For each subnetwork, we calculate the mean squared error (MSE) between the output electric-field intensity $I_k^{M+1}$ and the corresponding target $T_k$ as the loss function:

$$L(\varphi_k) = \frac{1}{N/4} \sum_1^{N/4} (I_k^{M+1} - T_k) \quad (9)$$

Meanwhile, the loss function of the full network can be defined as follows:

$$L(G) = \frac{1}{N} \sum_1^N (I^{M+1} - T) \quad (10)$$

Subsequently, by adding all the loss functions up, the phase parameters of all modules can be updated through error backpropagation and the gradient descent algorithm.

Moreover, by assembling the subnetworks after rotating them at different angles ($r, r = 90°, 180°, 270°$), we can obtain new networks with distinct functionalities. The parameters of such a network are denoted by $G^r$, which can be defined as:

$$G^r[:,0:N/2,0:N/2\,] = \varphi_1^r \tag{11}$$
$$G^r[:,0:N/2,N/2:N\,] = \varphi_2^r \tag{12}$$
$$G^r[:,0:,N/2:N,0:N/2\,] = \varphi_3^r \tag{13}$$
$$G^r[:,0:N/2:N,N/2:N\,] = \varphi_4^r \tag{14}$$

Similarly, the optimization of parameters can be accomplished by simply adding the loss function of the newly constructed network to those of the previous tasks and performing backpropagation jointly.

## 2.2 Design and fabrication

In this work, the wavelength of the incident light is set as λ = 1030 μm (0.29 THz), and the lateral size of each neuron (pixel) on the diffractive layer is 800 μm. During the training process, each pixel on the diffractive layer was meshed into a 4×4 grid to improve the simulation accuracy of the scalar diffraction model. Each PDNN model in this work was trained using Python (v3.12.4) and PyTorch (v 2.1.0) on a Windows 11 operating system (Microsoft) with AMD Ryzen 9 7950X 16-Core central processing unit, 128 GB of RAM, and a GeForce RTX 3090 graphical processing unit.

After training, we fabricated the diffractive layers using 3D printing. The material used in our fabrication is commercially available HTL resin with a refractive index n = 1.733 + 0.03j. First, we converted neurons' phase values ($\phi$) into a relative height map (Δz) using the following equation:

$$\Delta z = \frac{\lambda \phi}{2\pi \Delta n} \tag{15}$$

where Δn is the refractive index difference between the 3D printing material and air, and in this work Δn = 0.733. Accordingly, to achieve a 2π phase shift at 0.29 THz, the height of each pixel in the diffractive layers ranges from 0 to 1400 μm. Meanwhile, due to the limited resolution of 3D printing, the pixel heights were quantized with a precision of 10 μm, resulting in 1400/10+1=141 discrete height levels. In addition, to enhance the stiffness and stability of the diffractive layer, a 1 cm-thick substrate was added during the printing process. The optical setup for experimental characterization of PDNN is shown in Note S2 (Supporting Information).

## 3. Results

## 3.1 PDNN-enabled multi-degree-of-freedom terahertz holograms

To achieve the multi-degree-of-freedom terahertz holographic imaging, we designed a PDNN incorporating a diffractive layer consisting of 60×60 pixels, resulting in an x–y plane dimension size of 48 mm × 48 mm. Each quadrant in the PDNN corresponds to a subnetwork of 30×30 pixels, which are designed to generate holographic images of digits "0", "1", "2", and "5", respectively, and the fully assembled network is designed to generate holographic image of the digit "7". Fig. 2a shows the loss evolution for each network during the training process. A total of 500 epochs is adopted during training, and it can be observed that the networks converged rapidly after the first few training iterations. The optimized pixel phase distributions of the proposed PDNN are shown in Fig. 2b, and the optical image of the fabricated diffractive layer is shown in Fig. 2c. Each quadrant of the diffractive layer corresponds to one subnetwork. When the functionality of a specific subnetwork is required, a metallic mask (Fig. S1) is employed to block light in other regions while retaining illumination only in the area corresponding to the target subnetwork, thereby achieving its selective activation. When the full functionality of the entire network is needed, removing the mask allows all neurons within the network to be activated simultaneously. For each subnetwork, the distances from the mask to the diffractive layer and from the diffractive layer to the output plane are set to 10 mm and 30 mm, respectively. For the assembled network, the distances from the mask to the diffractive layer and from the diffractive layer to the output plane are set to 10 mm and 40 mm.

Fig. 3a shows the target images for the multi-degree-of-freedom terahertz holographic imaging and Fig. 3b demonstrates the corresponding simulation results based on the angular spectrum diffraction theory. It can be clearly observed that digits "0", "1" "2", "5", and "7" are projected on the predefined output plane. To further verify the optimization results obtained from scalar diffraction theory, we imported the STL model of the diffractive layer into LUMERICAL software and performed more precise full-wave 3D finite difference time-domain (FDTD) simulations. The incident light was set as a plane wave, the boundary condition was defined as perfectly matched layer (PML), and the mesh accuracy was set to 3. The FDTD simulation results are shown in Fig. 3c, which match well with the results obtained from the angular spectrum diffraction.

The experimentally measured normalized light intensity distributions at the output plane for different networks are shown in Fig. 3d. The experimental results validate the successful operation of all functional modules. We further calculated the diffraction efficiency for each terahertz hologram, defined as the ratio of the total optical intensity within the target pattern region to the total input optical intensity. The diffraction efficiencies derived from angular spectrum diffraction simulations and experimental measurements are as follows: "0" (48.17% for simulation and 16.32% for experiment), "1" (48.44% for simulation and 14.53% for experiment), "2" (46.98% for simulation and 16.33% for experiment), "5" (48.12% for simulation and 16.65% for experiment) and "7" (39.21% for simulation and 13.49% for experiment). The reduction in experimental efficiency is primarily caused by the absorption and reflection of light within the diffractive layer.

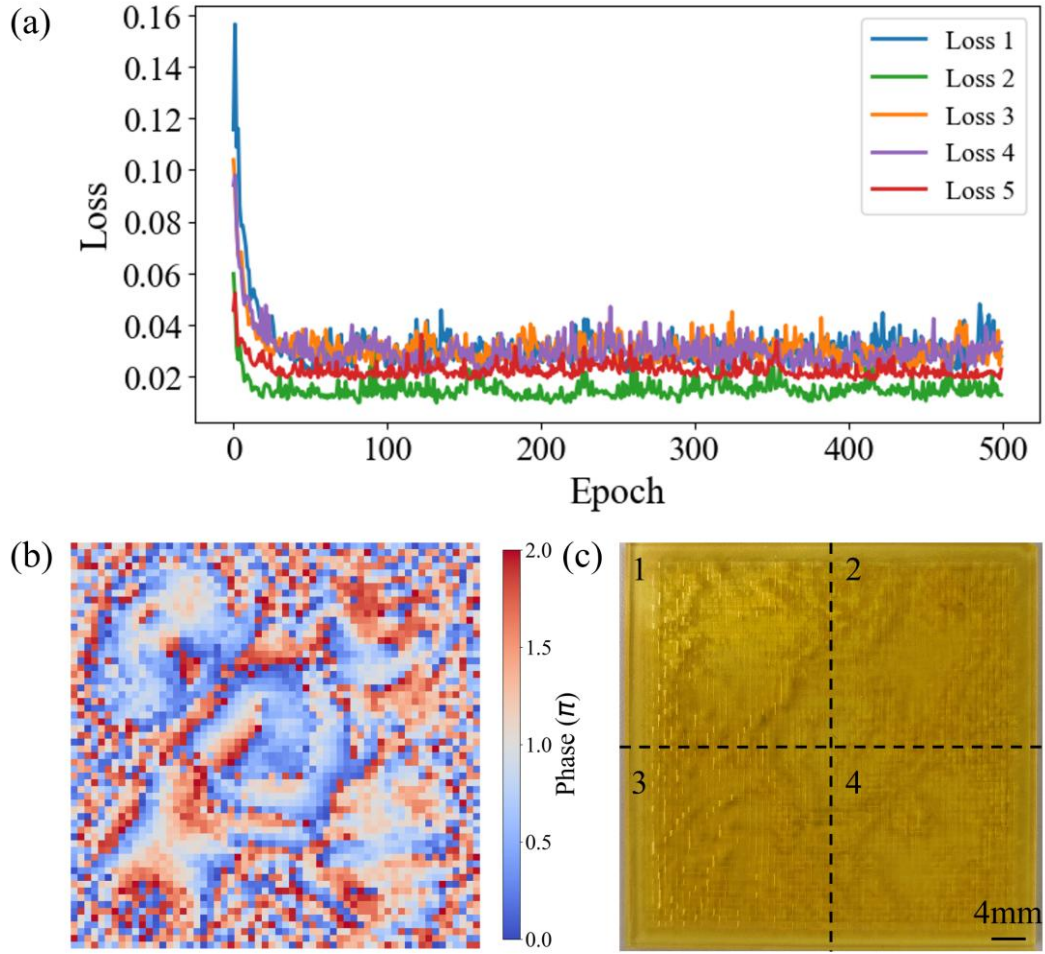

Fig. 2. (a) Evolution of training loss during the 500-epochs training process; (b) Optimized pixel phase distribution for the diffractive layer; (c) Optical image of the 3D printed diffractive layer, where each quadrant corresponds to a subnetwork designed to generate holographic images of digits "0", "1", "2", and "5", respectively, and the fully assembled network is designed to generate holographic image of the digit "7".

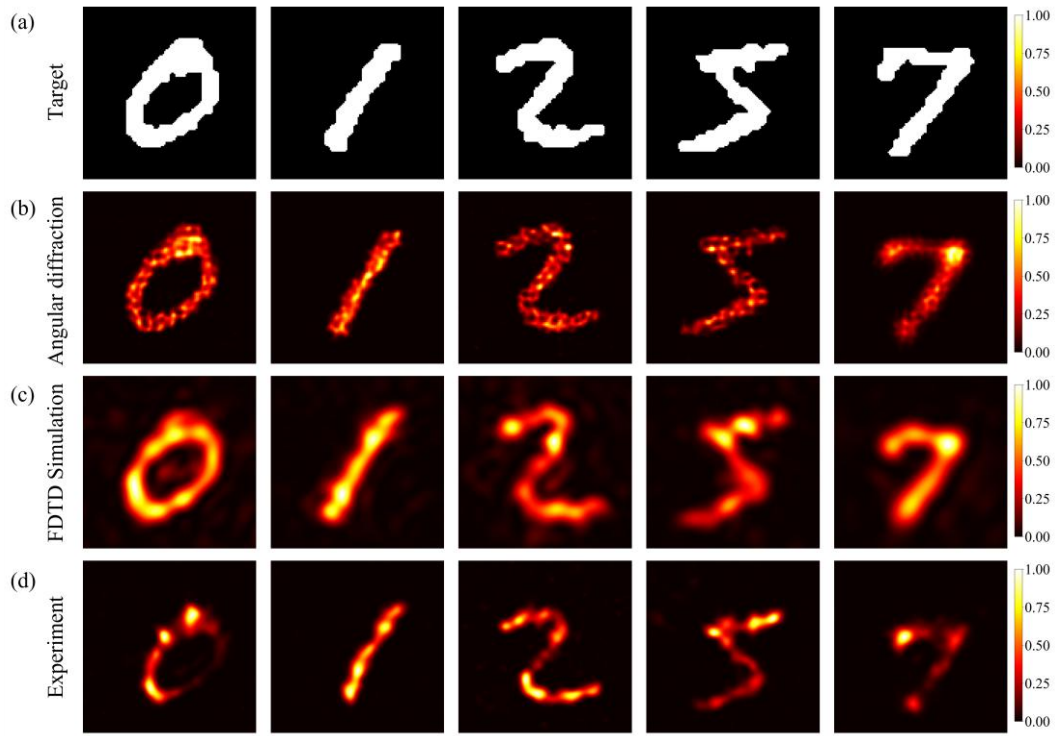

Fig. 3. (a) Target patterns for multi-degree-of-freedom terahertz holographic imaging (digits "0", "1", "2", "5", and "7"); (b) Normalized hologram intensity distribution obtained from angular diffraction; (c) Corresponding FDTD simulation results; (d) Experimentally measured diffraction field intensity.

## 3.2 PDNN-enabled terahertz holographic imaging and handwritten digit classification

We further verified the feasibility of integrating two different tasks—holographic imaging and handwritten digit classification—within the PDNN. Similar to the design in Section 3.1, the PDNN comprises a diffractive layer with 60×60 pixels. The four subnetworks are designed to generate holographic images of the letters "S", "J", "T", and "U", respectively, while the fully assembled network is trained to classify handwritten digits "1" and "3". For each subnetwork, the distances from the mask to the diffractive layer and from the diffractive layer to the output plane are set to 10 mm and 30 mm, respectively, while for the complete network, these two distances are set to 20 mm and 40 mm.

Fig. 4a shows the evolution of the loss values of each subnetwork during training. The training loss decreases rapidly in the initial stage and begins to saturate after about 1000 epochs. The optimized phase distribution of the PDNN and the optical image of the 3D-printed diffractive layer are shown in Fig. 4b. The simulation and experimental results for each holographic imaging task are demonstrated in Fig. 4c. Similar to section 3.1, simulations were performed using both the angular spectrum diffraction theory and the FDTD method, and both sets of results agree well with the experimental measurements. In addition, we also calculated the diffraction efficiencies of the subnetworks in holographic imaging through angular spectrum diffraction simulation and experimental measurement. The calculated efficiencies for each scenario are: "S" (47.52% for simulation and 28.27% for experiment), "J" (48.17% for simulation and 26.15% for experiment), "T" (47.05% for simulation and 27.17% for experiment), and "U" (48.19% for simulation and 25.83% for experiment), respectively.

For the classification task, the classical dataset from the Modified National Institute of Standards and Technology (MNIST) is utilized to train the recognition task of handwritten digits (digits "1" and "3"). The convergence curves during training for classification tasks are provided in Fig. 5a. Confusion matrices summarizing the classification performance are provided in Fig. 5b and 5c, showing a recognition accuracy of 95.16% in simulation and 100% in experimental measurements. The simulation and experimental validation results are illustrated in Fig. 5d-g. To mimic the input of handwritten digit images, the input plane is constructed using a metallic mask containing the handwritten digit pattern (Fig. 5d). When the plane wave passes through the mask, it carries the amplitude information of the object to be recognized, which is then processed by the diffractive layer and focused onto specific detection regions on the output plane. For the incidence of the digit "1" and "3", the diffractive field is modulated by the diffractive layer and finally focused onto the left-center and right-center recognition regions on the output plane, respectively. The calculated and measured energy distributions are illustrated in Fig. 5h and 5i, demonstrating that PDNN can correctly recognize the handwritten digits according to the energy distribution.

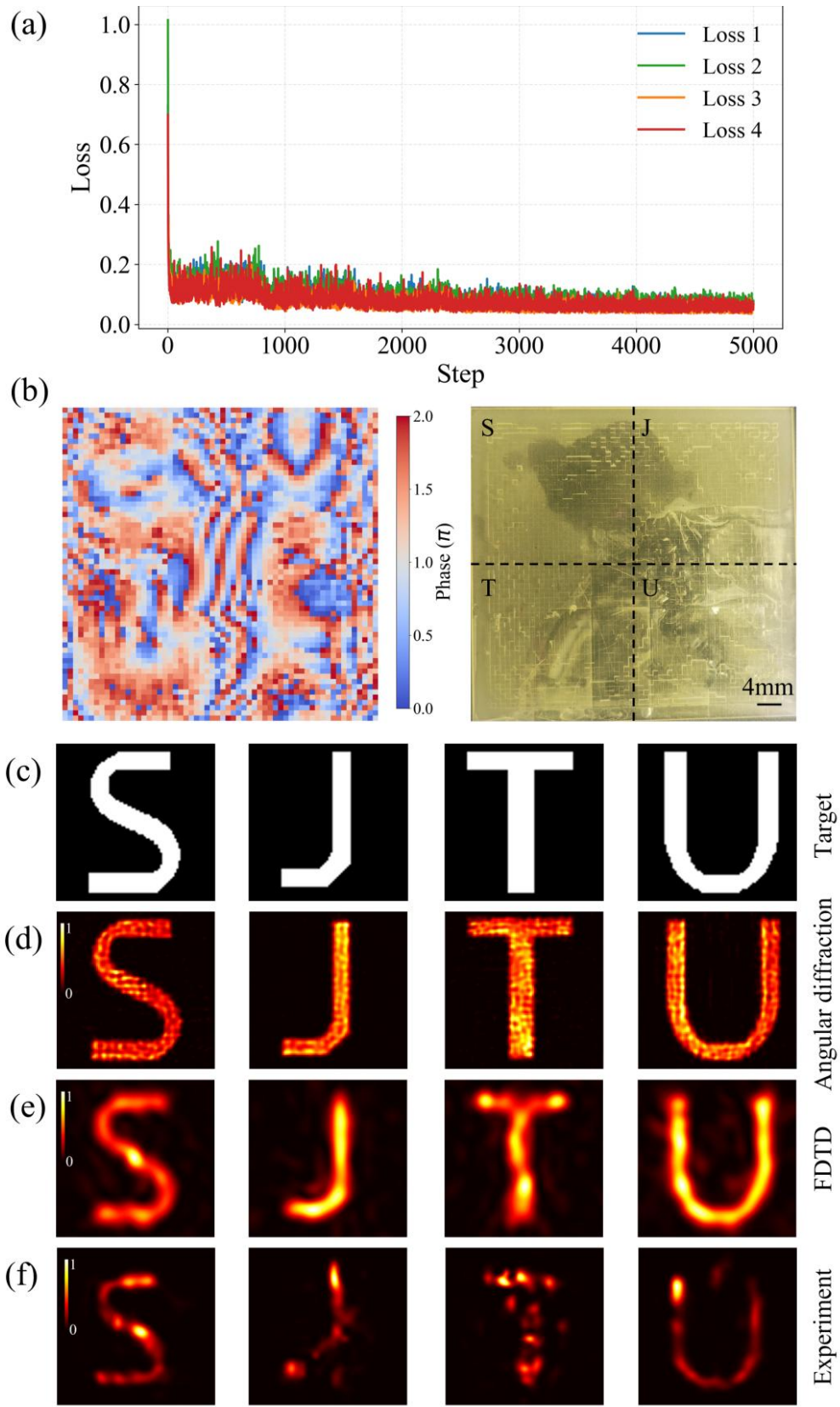

Fig. 4. (a) Training loss evolution for the holographic imaging tasks; (b) Optimized pixel phase distribution for the diffractive layer and optical image of the 3D printed diffractive layer; (c) Target patterns for holographic imaging (letters "S", "J", "T", and "U") ;(d) Normalized hologram intensity distribution obtained from angular diffraction; (e) Corresponding FDTD simulation results; (f)

Experimentally measured diffraction field intensity.

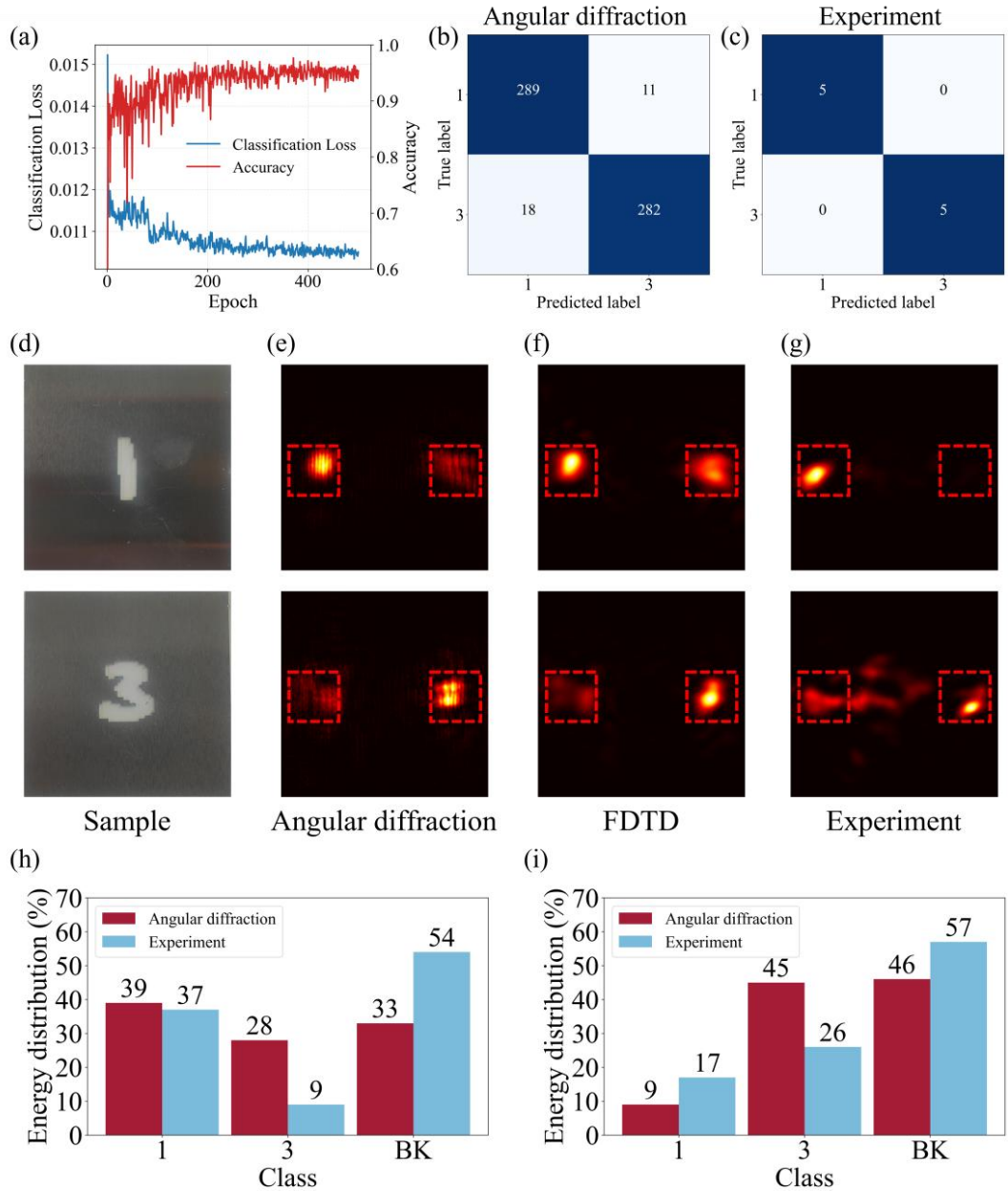

Fig. 5. (a) Evolution of training loss and accuracy during 500 epochs training process; (b, c) Simulation and experimental results of confusion matrix for handwritten digits; (d) Optical images of the fabricated handwritten digit samples; (e)-(g) Output-field intensity-distributions obtained from (e) angular diffraction simulation, (f) FDTD simulation, and (g) experimental measurement; (h, i) Calculated and experimentally measured energy distribution of handwritten digits (h) "1" and (i) "3" classification, where 'BK' denotes the background region.

## 3.3 Enhancing task integration in PDNNs via subnetwork rotation

Finally, we demonstrated that the PDNN can generate new networks with distant functionalities by combining subnetworks after rotating them at various angles. As shown in Fig. 6, the four subnetworks are designed to generate holographic images of the digits "1", "2", "3", and "4", respectively. Each subnetwork was then rotated clockwise around the z-axis by 0°, 90°, 180°, and 270°, respectively, and combined laterally to form different new networks, each capable of generating the holographic image of digits "5", "6", "7", and "8". The loss convergence curves of the networks during the training phase are shown in Fig. 7a. The optimized phase distributions are shown in Fig. 7b.

After training, the performance of the designed PDNN was verified using both the angular spectrum method and FDTD simulations, as illustrated in Fig. 7c. The results clearly show that that the designed networks successfully generate the desired patterns on the output planes in all scenarios, and the results of the FDTD simulations are consistent with those obtained from the angular spectrum simulations. And the diffraction efficiencies calculated from the angular spectrum diffraction simulations for each scenario are: "1" (43.94%), "2" (44.29%), "3" (44.41%), "4" (44.87%), "5" (45.09%) "6" (45.31%), "7" (45.49%), and "8" (45.07%).

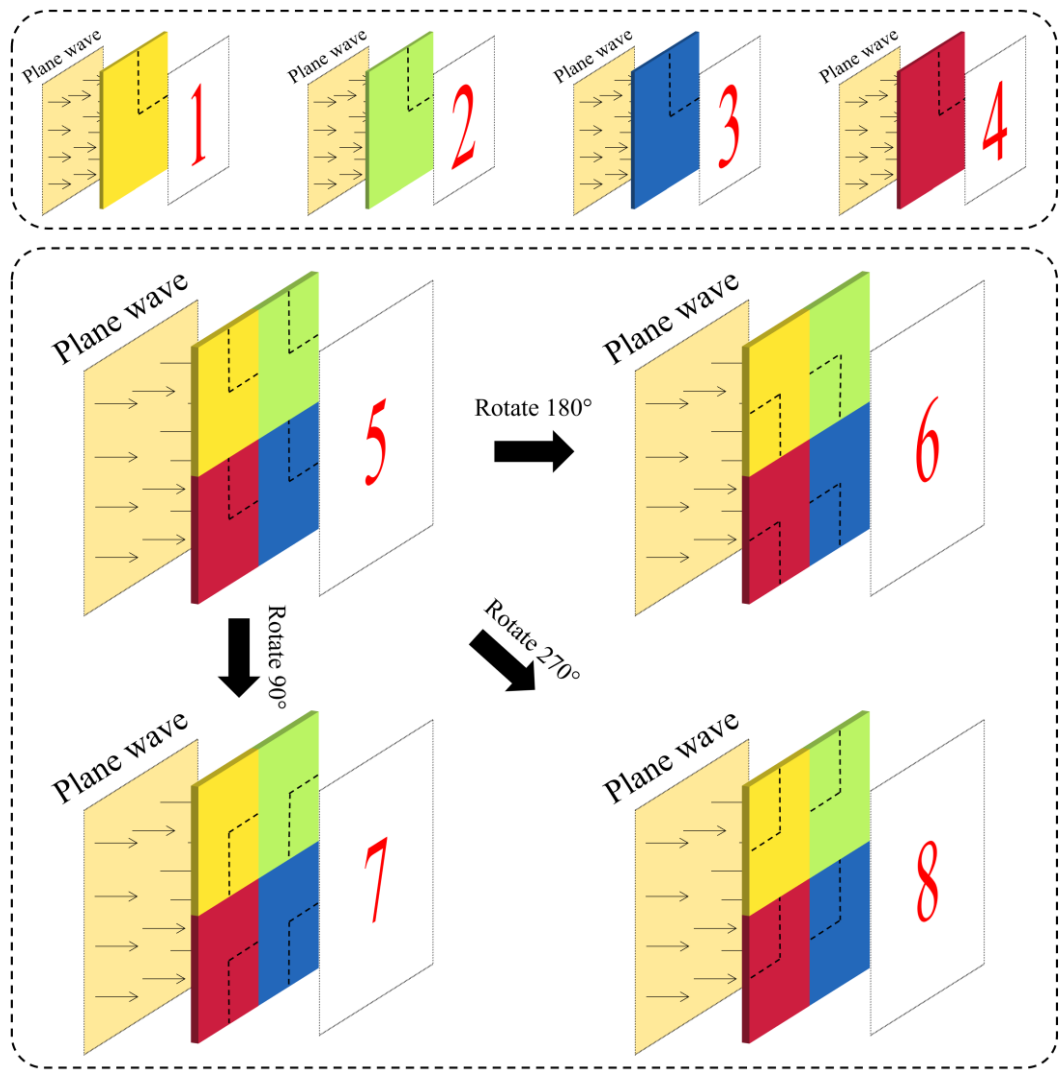

Fig. 6. Each sub-network is individually designed to generate holograms of the digits "1", "2", "3", and "4". Furthermore, a composite network formed by rotating each sub-network clockwise about the z-axis by 0°, 90°, 180°, and 270° can be used to generate holographic patterns corresponding to the digits "5", "6", "7", and "8".

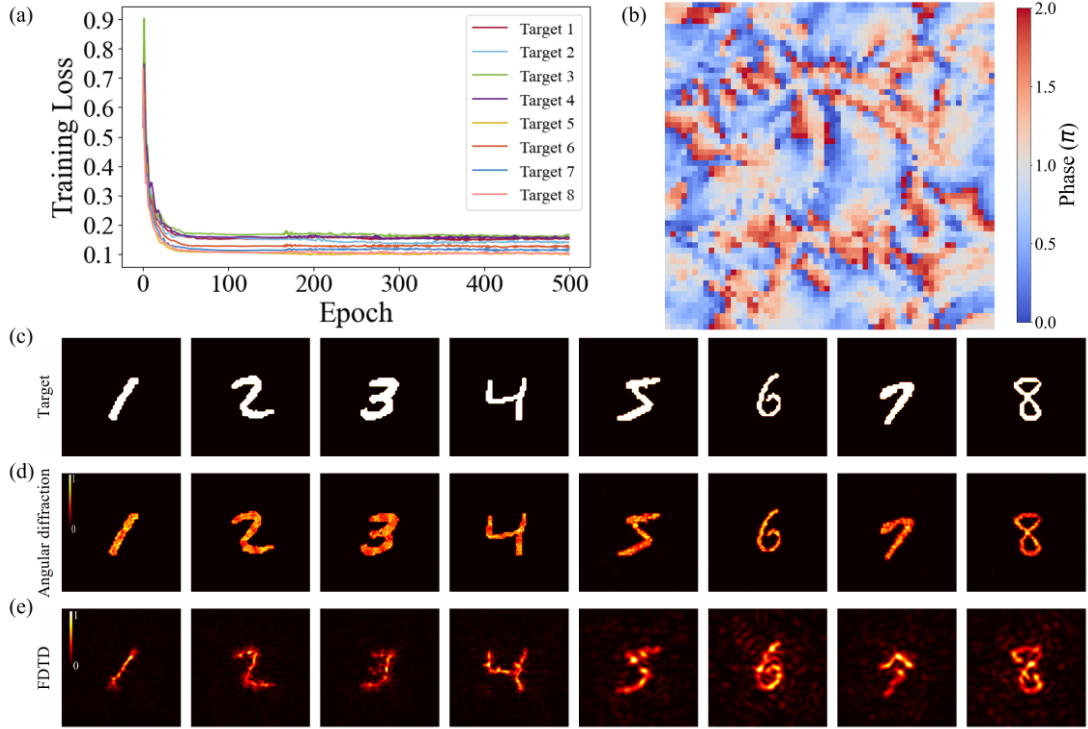

Fig. 7. (a) Evolution of training loss during 500 epochs training process; (b) Optimized pixel phase distribution for the diffractive layer; (c) Target patterns for holographic imaging (digits "1" - "8"); (d) Normalized output electric-field intensity distributions obtained from angular diffraction; (e) Corresponding FDTD simulation results.

## 4. Conclusion

In this work, we proposed a horizontally scalable multi-task diffractive neural network framework named PDNN, and validated its effectiveness through both simulations and experiments. By horizontally combining four subnetworks with independent functionalities, a new network with additional capabilities is generated, enabling the entire system to simultaneously perform five different functions. Moreover, the PDNN supports assembling subnetworks after rotating them by different angles, where each rotation configuration leads to a newly formed network with distinct functionality, thereby greatly enhancing the overall task integration capacity of the system. The simulation and experimental results are both consistent with the expected outcomes, validating the proposed approach. In addition, as a simple and general framework, the PDNN can be further extended by incorporating other multiplexing strategies during its design, such as wavelength multiplexing, polarization multiplexing, and orbital angular momentum multiplexing. The PDNN framework effectively addresses the limited extensibility of traditional DNNs. By demonstrating a pathway toward hardware-reconfigurable, multi-task optical inference, this work opens up new possibilities for the development of compact and versatile all-optical intelligence systems.

## Acknowledgements

This research was sponsored by the National Natural Science Foundation of China (62375170, 62535019, 62235004, and 62522509), the Shanghai Jiao Tong University (YG2024QNA51), and